\documentstyle[12pt]{article}
\begin{document}
\rightline{hep-th/9606174, VPI-IPPAP-96-3}
\vskip 0.2in
\centerline{\Large Invariant regularization of anomaly-free chiral theories}
\vskip 0.2in
\centerline{Lay Nam Chang ${}^*$}
\vskip 0.10in
\centerline{and}
\vskip 0.10in
\centerline{Chopin Soo ${}^\dagger$}
\vskip 0.2in
\centerline{Department of Physics, and}
\centerline{Institute for Particle Physics and Astrophysics,}
\centerline{Virginia Tech,}
\centerline{Blacksburg, VA 24061-0435, U.S.A.}
\vskip 0.20in
\centerline{PACS number(s): 11.15.-q, 11.30.Rd, 04.62.$+$v}
\vskip 0.10in

We present a generalization of the Frolov-Slavnov invariant 
regularization scheme for chiral fermion theories in curved spacetimes. 
The Lagrangian level regularization is explicitly invariant under all the 
local gauge symmetries of the theory, including local Lorentz invariance. 
The perturbative scheme works for {\it arbitrary} representations which 
satisfy the chiral gauge anomaly and the mixed Lorentz-gauge anomaly 
cancellation conditions. Anomalous theories on the other hand manifest 
themselves by having divergent fermion loops which remain unregularized by 
the scheme. Since the invariant scheme is promoted to also include local 
Lorentz invariance, spectator fields which do not couple to gravity cannot be, 
and are not, introduced. Furthermore, the scheme is truly chiral (Weyl) in 
that {\it all} fields, including the regulators, are left-handed; and 
{\it only the left-handed spin connection} is needed. The scheme 
is, therefore, well suited for the study of the interaction of matter 
with all four known forces in a completely chiral fashion.   
In contrast with the vectorlike formulation,
the degeneracy between the Adler-Bell-Jackiw current and the fermion number
current in the bare action is preserved by the chiral regularization scheme. 

\vfil
$^\dagger$Electronic address: soo@theory.uwinnipeg.ca. Present address:Dept. 
of Physics, University of Winnipeg, Winnipeg, Manitoba, Canada R3C 2E9. 
$^*$Electronic address: laynam@vt.edu \hfil

%
%
%
\bigskip
\section*{I. INTRODUCTION}
\bigskip


  It is believed that the existence of an invariant regularization 
for a quantum field theory of chiral fermions is predicated upon  
the absence of perturbative anomalies.\footnote{This does not  
necessarily mean that some version of the theory cannot be 
defined. For instance, the anomalous chiral Schwinger model can be solved 
exactly. However, the gauge invariance of the theory is lost.} 
It is interesting to ask whether an explicitly invariant regularization scheme 
for chiral fermions can be constructed so that it successfully regularizes 
the theory when the representation of the chiral fermion multiplet is 
anomaly free, and fails to do so precisely when the representation is not.
In the following, we present a scheme of
regularization at the Lagrangian level incorporating
this feature, which is suitable for 
describing
chiral theories in curved spacetimes. 

Frolov and Slavnov\cite{FS} first proposed an explicitly gauge invariant 
regularization which makes essential use of an infinite tower of 
Pauli-Villars-Gupta regulators\cite{Pauli}.
The theory was originally based on the SO(10) multiplet\cite{Georgi} - or 
rather, the 
16-dimensional chiral representation of 
Spin(10) - with the Standard Model embedded in it.\footnote{Chirality in
Spin(10) is defined relative to $\Gamma_{11}$, which is proportional to
the product of all ten Dirac matrices spanning the Clifford algebra in
ten dimensional Euclidean space.} 
Since 
invariant regulator mass terms are required for this type of 
Pauli-Villars-Gupta regularization, it is necessary to ``double'' in 
internal gauge group space 
by also including fields which transform according to the complex conjugate 
representation\cite{Aoki}.
The tower of regulators is therefore neutral with respect 
to $\Gamma_{11}$, and can regularize only the 
singlet part
of the $\frac{1}{4}(1-\Gamma_{11})(1-\gamma^5)$ projection of the bare 
gauge current. 
However, it is shown that the $\Gamma_{11}$ part gives rise to no further 
divergences
due to the fact that the trace of four or less generators of Spin(10) 
with $\Gamma_{11}$ vanishes. Frolov and Slavnov also proposed a 
discretization of the theory{\cite{FS2}. It is believed that the 
Nielsen-Ninomiya no-go 
theorem\cite{Nielsen} is surmounted by the presence of the {\it infinite} 
tower of regulator fields. 
It was not immediately  clear from this discussion  if the method 
generalizes to arbitrary 
anomaly-free chiral theories, although 
it was clear how to regularize theories based upon  SO($2n\geq 10$) groups. 
Okuyama and 
Suzuki\cite {Okuyama} clarified and generalized the original 
Frolov-Slavnov idea to include fermion multiplets in arbitrary real and 
pseudoreal representations.  But the 
generalization to curved spacetimes and Abelian gauge groups remained somewhat 
unclear. 

The scheme was considered in a different light by Fujikawa\cite{Fujinew}, and 
by Narayanan and Neuberger\cite{Nara}. They doubled in external or 
Lorentz space by including right-handed as well as left-handed 
regulator fields.
In these vectorlike formulations, no doubling in the internal symmetry group
is needed.
To study nonperturbative effects, Narayanan and Neuberger\cite{Naraneu} also 
proposed the ``overlap formalism'' by treating the extra 
index associated with the tower of regulators as an additional dimension of 
spacetime, and by defining the chiral fermion determinant as the
overlap of two different gound states of the higher dimensional theory.  
As such, the Nielsen-Ninomiya no-go 
theorem of putting chiral fermions on a lattice may be overcome by treating
the chiral theory in 2$n$ dimensions as the target of another in 2$n+1$ 
dimensions in which there is no concept of chirality.

In these vectorlike formulations, the tower is parity even or 
$\gamma^5$ neutral. So doubling in external space by including
right-handed fields now regularizes the left-right symmetric part, 
 leaving the $\gamma^5$ part of the bare gauge current untouched. For 
anomaly-free theories, it can be argued that there are no divergent 
parity-odd diagrams generated by the  gauge current.  As a result, the 
{\it gauge} current is regularized for
arbitrary, perturbatively anomaly-free representations. 
However, as noted by Fujikawa\cite{Fujinew}, the parity-nonconserving 
amplitude with gauge singlet currents can be divergent. For instance, 
the fermion number current is not free of divergences, and is different 
from the axial or Adler-Bell-Jackiw(ABJ) current at the
regularized level\cite{Fujinew}. 
On the other hand, in the totally left-handed 
Frolov-Slavnov formulation, the degeneracy between the well-defined 
ABJ singlet current and the fermion number current in the
original bare action is preserved by the chiral regularization scheme.

What happens when the Frolov-Slavnov regularization scheme is promoted to 
include {\it local} Lorentz invariance and the effects of curved spacetimes? 
Two issues immediately arise.   
First, it is known that chiral fermions in curved spacetimes can introduce a
further perturbative mixed Lorentz-gauge anomaly\cite{nieh}.
Second, all fields couple to
gravity, and the trick of introducing spectators inherent in some
methods of defining chiral theories may not work\cite{Okuyama,ap}.
For example, 
in Ref\cite{FS}, the Standard Model (or any anomaly-free 
subgroup of SO(10)) can be recovered by taking the gauge 
field $W_\mu$ to lie 
only in the relevant subgroup. However, if the regularization is 
extended to include invariance under local Lorentz transformations, then
the extra spectator ``neutrino,'' which is not coupled 
to any internal gauge field, becomes {\it physical} as a result of 
its coupling to gravity.  

Similar remarks apply to right-handed fermions. These make an appearance
in a chiral theory either as regulators, as in vectorlike schemes, or
as spectators in defining propagators\cite{Okuyama}. 
All these fields get coupled to gravity and 
become physically interacting degrees of freedom.\footnote{It can be shown 
later on that the 
regularized gauge current in curved spacetimes does
reduce to the original current as the regulator masses go to infinity.} 
The key point is that there can be {\it no} passive spectators if 
the regularization scheme is promoted to also respect local Lorentz 
invariance.   
 
There is yet another issue we need to be aware of.
Right-handed multiplets can be introduced
in a covariant way for curved spacetimes only if one also allows for 
right-handed spin connections.   
It is known through the work of Ashtekar and 
others\cite{ash, samuel} that the (anti-)self-dual formulation of gravity 
which involves only the left-handed spin connection, 
rather than the full spin connection, may provide a complete description 
of gravity in four dimensions. These right-handed spin connections 
are generally not independent of the left-handed ones\cite{ash}, and their
presence might therefore complicate the gravity field equations 
unnecessarily.\footnote{Recall that even a Majorana fermion couples to both 
the left- and right-handed spin connections.}
Indeed in the (anti-)self-dual formulation, no right-handed fermions should 
be introduced\cite{ART, cps}. 

For these reasons, we examine in this paper a regularization scheme
that is based only upon  left-handed fields, with no spectators.
This scheme extends the Frolov-Slavnov
regularization to anomaly-free chiral theories in arbitrary 
complex representations 
 in curved spacetimes.\footnote{The representations may be 
reducible, as in the SU(5) grand unified theory (GUT) model.} 
The regularized chiral fermion action is explicitly gauge, Lorentz, and 
diffeomorphism invariant, and is truly chiral (Weyl) in the sense that 
only left-handed spin connections and left-handed multiplets 
are introduced. 
The proposed regularization is therefore well 
suited for the study of the interaction of matter with all the four 
known forces in a completely chiral manner\cite{cps}.

We shall show that the generalization regularizes the chiral
theory if and only if the theory is free of all perturbative chiral gauge 
anomalies, including the Lorentz-gauge mixed anomaly. In this explicitly 
invariant scheme, anomalous theories manifest themselves by having divergent
fermion loops which remain unregularized.

\bigskip
\section*{II. BARE ACTION AND INVARIANT MASSES}
\bigskip
  
The bare chiral fermion action may be taken to be
\begin{equation}
{\cal S}_{F_{bare}} =\int d^4x e{\overline{\Psi}}^-_{L_0}i
{{D\kern-0.15em\raise0.17ex\llap{/}\kern0.15em\relax}}P_L
\Psi^-_{L_0},
\end{equation}
where $i{D\kern-0.15em\raise0.17ex\llap{/}\kern0.15em\relax} =\gamma^\mu
(i\partial_\mu + W_{\mu a}T^a + \frac{i}{2}
 A_{\mu AB}{\sigma}^{AB})$, ${\sigma}^{AB}={1\over 4}
[\gamma^A, \gamma^B]$, and $e$ denotes the determinant of the vierbein.
$P_L = \frac{1}{2}(1 - \gamma^5)$ is the left-handed projection operator. 
We adopt the convention
\begin{equation}
\{ \gamma^{A}, \gamma^{B}\}= 2\eta^{AB},
\end{equation}
with $\eta^{AB} = {\rm diag}(-1,+1,+1,+1)$. Lorentz indices are denoted by 
uppercase Latin indices while Greek indices are spacetime indices.

In general, the fermion multiplet $\Psi^-_{L_0}$ is in a complex 
representation. 
Recall that if the generators $T^a$ satisfy
\begin{equation}
\left[T^a, T^b \right] = if^{ab}\,_cT^c,
\end{equation}
then $(-T^a)^*$ satisfy the same Lie algebra.\footnote{We adopt the convention 
of $(T^a)^\dagger = T^a$ and real structure constants. For invariance of 
the action, the representation has to be unitary.} 
If there exists an $S$ such 
that $S^{-1}(-T^a)^* S = T^a$, then the representation is called 
real (pseudoreal) if $S$ is symmetric (antisymmetric). Otherwise, the 
representation is termed complex.

With only a single left-handed multiplet, Lorentz-invariant mass terms are 
Majorana in nature. The simple form of 
$\Psi^T_L C_4 \Psi_L$ is not invariant under internal symmetry 
transformations.\footnote{$C_4$ is the charge conjugation matrix in four 
dimensions with $C^T_4 = C^{-1}_4= C^{\dagger}_4 = -C_4$.} 
However, with real representations, an invariant mass term 
$m\Psi^T_LSC_4 \Psi_L $ can be 
constructed from a {\it single} multiplet. 
Observe that for a nonvanishing mass term, $S$  has to be symmetric for 
anticommuting fields and antisymmetric for commuting fields.
For complex representations, a gauge and Lorentz 
invariant mass term cannot be made out of a single multiplet. This poses a
challenge for the usual invariant Pauli-Villars-Gupta regularization, even 
though the chiral fermions may belong to an anomaly-free representation.  

We shall generalize the method of Frolov and Slavnov by including regulators
which are doubled in internal space.  This doubling is achieved by
including fermions which transform
according to the $(-T^a)^*$ representation, and then
an invariant mass term can be
formed because under
\begin{equation}
\Psi^-_{L_r} \rightarrow e^{i\alpha_a T^a}\Psi^-_{L_r}, \qquad
\Psi^+_{L_r} \rightarrow e^{i\alpha_a(-T^a)^*}\Psi^+_{L_r},
\end{equation}
the combination
$\left[(\Psi^+_{L_r})^TC_4 \Psi^-_{L_r} + (\Psi^-_{L_r})^TC_4 \Psi^+_{L_r} 
 + H.c.\right]$ is 
invariant under internal gauge and Lorentz transformations.

We introduce in the enlarged space the quantities
\begin{equation}
{\cal T}^a \equiv \left(\matrix{(-T^a)^* &0\cr 0& T^a}\right),\qquad 
\sigma^1 \equiv \left(\matrix{0 & 1_d\cr 1_d &0}\right),\qquad
\sigma^3 \equiv \left(\matrix{1_d & 0\cr 0 &-1_d}\right),
\end{equation}
where $d$ denotes the number of internal components of the original 
multiplet.  In this notation,  the original multiplet can be expressed as
\begin{equation}
\left[\matrix{0\cr \Psi^-_{L_0}}\right]={1\over2}(1_{2d}-\sigma^3)\Psi_{L_0},
\end{equation}
 and the mass terms for the regulator fermions, 
\begin{equation}
\Psi_{L_r} = \left[\matrix{\Psi^+_{L_r} \cr \Psi^-_{L_r}}\right],
\end{equation}
can be written as $m_r(\Psi^T_{L_r}\sigma^1C_4\Psi_{L_r} + H.c.).$ 
The doubled regulator fermion multiplets are to be coupled to the 
2$d$-dimensional representation of the gauge connection, $W_{\mu a}{\cal T}^a$.

The $\Psi_{L_r}$ fields are assumed to be anticommuting. Commuting doubled 
regulator fields $\Phi_{L_s}$ are introduced in a similar manner. 
These have mass terms
\begin{equation}
m_s\Phi^T_{L_s}(-i\sigma^2)C_4\Phi_{L_s}= m_s\left[-(\Phi^+_{L_s})^T
C_4\Phi^-_{L_s} +  (\Phi^-_{L_s})^TC_4\Phi^+_{L_s}\right],
\end{equation}
with
\begin{equation}
-i\sigma^2 \equiv \left(\matrix{0 & -1_d\cr 1_d & 0}\right)= \sigma^1\sigma^3.
\end{equation}
Note that these invariant mass terms for the doubled anticommuting and 
commuting fields exist, because for the ${\cal T}^a$ representation, there
is a symmetric ($\sigma^1$) and an antisymmetric $(-i\sigma^2)$ matrix 
which satisfy\footnote{A set of conditions for generalized Pauli-Villars
regularization is also given in \cite{Okuyama}.}
\begin{equation}
(\sigma^1){\cal T}^a (\sigma^1)^{-1}=(-i\sigma^2){\cal T}^a (-i\sigma^2)^{-1} 
= (-{\cal T}^a)^* .
\end{equation}
It is clear that these constructs work for arbitrary groups and 
representations $T^a$. Note also that all the fields 
are left-handed.

We shall next show that this generalization of the Pauli-Villars-Gupta method
can regularize chiral fermions perturbatively in the original sense of 
Frolov and Slavnov if and only if the conditions for perturbative anomaly
 cancellations, including the mixed Lorentz-gauge anomaly, are 
satisfied.   

\bigskip
\section*{III. REGULARIZED ACTION}
\bigskip

The total regularized action which is explicitly gauge and Lorentz and, also
diffeomorphism invariant 
is taken to be\footnote{We also allow all the fields to transform
under general coordinate transformations. Here, we regularize only
fermion loops in background fields, and do not address the question of 
the regularization of the gauge and gravitational fields. 
Gauge propagators may be regularized by other methods.
Full quantum gravity effects are beyond the scope of this paper.}
\begin{eqnarray}
{\cal S}_{F_{reg}}&=&{\int}d^4xe[{\sum_{r=0,2,...}}\{{\overline\Psi}_{L_r} 
i{{D\kern-0.15em\raise0.17ex\llap{/}\kern0.15em\relax}}\Psi_{L_r} 
+ {1\over 2}m_r(\Psi^T_{L_r}\sigma^1C_4\Psi_{L_r} 
+ {\overline\Psi}_{L_r}\sigma^1C^\dagger_4{\overline \Psi}^T_{L_r})\}\cr
&-&{\sum_{s=1,3,...}}\{{{\overline\Phi}_L}_s\sigma^3i
{{D\kern-0.15em\raise0.17ex\llap{/}\kern0.15em\relax}}\Phi_{L_s} + 
{1\over 2}m_s(\Phi^T_{L_s}\sigma^1\sigma^3C_4\Phi_{L_s} + 
{\overline\Phi}_{L_s}C^\dagger_4\sigma^3\sigma^1{\overline\Phi}^T_{L_s})\}]
\nonumber\\
\end{eqnarray}

The sums are over all even natural numbers for the anticommuting fields and 
over all odd natural numbers for the commuting fields. The usefulness of this 
convention will become apparent later on. With the exception of
\begin{equation} 
\Psi_{L_0} = {1\over 2}(1-\sigma^3)\Psi_{L_0}=
\left[\matrix{0\cr \Psi^-_{L_0}}\right] ,
\end{equation} 
which is the original and undoubled chiral {\it massless} 
($m_0 = 0$) fermion multiplet,
all other anticommuting $\Psi_{L_r}$ and commuting $\Phi_{L_s}$ multiplets 
are generalized Pauli-Villars-Gupta regulator fields, doubled in 
internal space, and endowed with Majorana masses, 
which we take for definiteness
to satisfy $m_n = n\Lambda$. We 
emphasize that due to the fact that all the multiplets are left-handed, 
there are no couplings to the right-handed spin connection which does 
not need to be introduced for the Weyl action.

In matrix notation, the regularized fermion action can be reexpressed as
\begin{eqnarray}
{\cal S}_{F_{reg}}&=&
\int dx \int dy {1\over 2}\{\sum_r 
\left[\matrix{{\tilde\Psi}^T_{L_r}(x) & 
{\tilde{\overline\Psi}}_{L_r}(x)}\right]
\left[\matrix{M_r(x,y)\sigma^1C_4& -i
{D\kern-0.15em\raise0.17ex\llap{/}\kern0.15em\relax}^T(x,y)\cr
i{{D\kern-0.15em\raise0.17ex\llap{/}\kern0.15em\relax}}(x,y)& M_r(x,y)
\sigma^1C^\dagger_4}\right]
\left[\matrix{{\tilde\Psi}_{L_r}(y)\cr
{\tilde{\overline\Psi}}^T_{L_r}(y)}\right]\cr
\nonumber\\
&-&\sum_s 
\left[\matrix{{\tilde\Phi}^T_{L_s}(x)& 
{\tilde{\overline\Phi}}_{L_s}(x)}\right]
\left[\matrix{M_s(x,y)\sigma^1\sigma^3C_4& i
{D\kern-0.15em\raise0.17ex\llap{/}\kern0.15em\relax}^T(x,y)\sigma^3\cr
\sigma^3i{{D\kern-0.15em\raise0.17ex\llap{/}\kern0.15em\relax}}(x,y)& 
M_s(x,y)C^\dagger_4\sigma^3\sigma^1}\right]
\left[\matrix{{\tilde\Phi}_{L_s}(y)\cr
{\tilde{\overline\Phi}}^T_{L_s}(y)}\right]\},
\end{eqnarray}
with
\begin{eqnarray}
M_n(x,y) &=& m_n\delta(x-y),\cr
\nonumber\\
i{D\kern-0.15em\raise0.17ex\llap{/}\kern0.15em\relax}(x,y)&=& e^{1\over 2}(x)
\gamma^{\mu}(x)
\left[(i\partial^x_\mu + W_{{\mu}a}(x){\cal T}^a +
{i\over2}A_{\mu{AB}}(x)\sigma^{AB})\delta(x-y)\right]e^{-{1\over2}}(y),\cr
\nonumber\\
i{{D\kern-0.15em\raise0.17ex\llap{/}\kern0.15em\relax}}^T(x,y) &=&
e^{-{1\over 2}}(x)\left[\{i\partial^y_\mu 
+ W_{{\mu}a}(y)({\cal T}^a)^T +{i\over 2}A_{\mu{AB}}(y)(\sigma^{AB})^T\}
\delta(y-x)\right]
{\gamma^\mu}^T(y)e^{1\over 2}(y).
\nonumber\\
\end{eqnarray}
To be compatible with the diffeomorphism-invariant measure\cite{Fujienergy},
\begin{equation}
\prod_{x,r}D[{\overline\Psi}_{L_r}(x)e^{1\over2}(x)]
D[e^{1\over2}(x)\Psi_{L_r}(x)],
\end{equation}
for the fields in curved spacetimes, we have also chosen to use
densitized variables defined by
\begin{eqnarray}
{\tilde{\overline\Psi}}_{L_r} &\equiv& {\overline\Psi}_{L_r} e^{1\over2}
\nonumber\\
{\tilde\Psi}_{L_r} &\equiv& e^{1\over2}{\Psi}_{L_r}, 
\end{eqnarray}
with a  similar set for 
${\tilde{\overline\Phi}}_{L_s}$ and ${\tilde\Phi}_{L_s}$.

For clarity, we shall use the explicit chiral 
representation 
\begin{eqnarray}
\gamma^{5}&=& \left(\matrix{1_{2}&0\cr 0&-1_{2}}\right), \cr
\nonumber\\
\gamma^{A}&= &\left(\matrix{ 0& i\tau^{A}\cr i{\overline \tau}^{A}&0}
\right).
\end{eqnarray}
In the above, $\tau^{a}= -{\overline \tau}^{a}$ ($a$=1,2,3) are Pauli
 matrices, and $\tau^{0} = {\overline \tau}^{0} = -I_{2}$. 

By writing in terms of left-handed two-component Weyl fermions,
\begin{equation}
{\Psi_L}_r = \left[\matrix{0\cr \psi_r}\right], \qquad 
{\Phi_L}_s=\left[\matrix{0\cr \phi_s}\right]
\end{equation}
and\footnote{For Lorentzian signature spacetimes, 
${\overline\psi}_{r} = \psi^\dagger_r\gamma^0$ whereas for Euclidean 
signature, ${\overline\psi}_{r} $  is treated as an independent field.}
\begin{equation}
\left[\matrix{{\overline\psi}_r& 0}\right]\equiv {{\overline\Psi}_L}_r,
\qquad
\left[\matrix{{\overline\phi}_s& 0}\right]\equiv {{\overline\Phi}_L}_s ;
\end{equation}
the propagators in background gauge and gravitational fields can then be 
read off as
\begin{eqnarray}
\langle T\{{\tilde\Psi}_{L_r}(x){\tilde{\overline\Psi}}_{L_r}(y)\} \rangle &=&
\left[\matrix{0 &0\cr
\langle T\{{\tilde\psi}_{L_r}(x){\tilde{\overline\psi}}_{L_r}(y)\}
\rangle & 0}\right]\cr
\nonumber\\
&=&-P_L e^{-{1\over2}}(iD_\mu)^\dagger{\gamma^\mu}e^{1\over2}
\frac{1}{m^2_r + e^{1\over2}\gamma^\mu iD_\mu e^{-1}
(iD_\nu)^\dagger\gamma^{\nu}e^{1\over2}}
\delta(x-y).
\nonumber\\
\end{eqnarray}
We have used the identities
\begin{eqnarray}
\sigma^1({\cal T}^a)^T(\sigma^1)^{-1}&=& -({\cal T}^a)^\dagger, \cr
\nonumber\\
C_4({{\sigma}^{AB}})^T(C_4)^{-1}&=& -{\sigma}^{AB}.
\end{eqnarray}

Furthermore, with respect to the Euclidean inner product 
$\langle {\tilde X}|{\tilde Y}\rangle = \int d^4x {\tilde X}^\dagger(x) 
{\tilde Y}(x)$, the Euclidean Dirac 
operator 
obeys\footnote{In writing down the propagator, we do not in general 
assume the absence of torsion. If the torsion vanishes, then
\begin{equation}
D_\mu e\gamma^{\mu} f =  e\gamma^{\mu}D_\mu f.
\end{equation}}
\begin{eqnarray}
(i{{\cal D}\kern-0.15em\raise0.17ex\llap{/}\kern0.15em\relax})^\dagger 
&\equiv&
(e^{1\over 2}{i{D\kern-0.15em\raise0.17ex\llap{/}\kern0.15em\relax}}
e^{-{1\over2}})^\dagger 
\nonumber\\
&=& (e^{1\over2}\gamma^{\mu}iD_\mu e^{-{1\over2}})^\dagger 
\nonumber\\
&=&e^{-{1\over2}}(iD_\mu)^\dagger\gamma^{\mu}e^{1\over 2},
\end{eqnarray}
where\footnote{The gauge fields $W_{{\mu}a}$ and $A_{\mu{AB}}$ are real and,
$\gamma^A$ and $i{\sigma}^{AB}$ are Hermitian for Euclidean signature 
manifolds.}
\begin{equation}
(iD_\mu)^\dagger = i\partial_\mu + W_{{\mu}a}{{\cal T}^a}^\dagger
+{i\over2}A_{\mu{AB}}{\sigma}^{AB}.
\end{equation}
The Euclidean propagator is therefore
\begin{equation}
\langle T\{{\tilde\Psi}_{L_r}(x){\tilde{\overline\Psi}}_{L_r}(y)\} \rangle
= -P_L (i{{\cal D}\kern-0.15em\raise0.17ex\llap{/}\kern0.15em\relax})^{\dagger}
\frac{1}{m^2_r + 
(i{{\cal D}\kern-0.15em\raise0.17ex\llap{/}\kern0.15em\relax})
(i{{\cal D}\kern-0.15em\raise0.17ex\llap{/}\kern0.15em\relax})^{\dagger}}
\delta(x-y).
\end{equation}
In a similar manner, for the commuting regulators,
\begin{eqnarray}
\langle T\{{\tilde\Phi}_{L_s}(x){\tilde{\overline\Phi}}_{L_s}(y)\} 
\rangle &=&\left[\matrix{0 &0\cr
\langle T\{{\tilde\phi}_{L_s}(x){\tilde{\overline\phi}}_{L_s}(y)\}
\rangle & 0}\right]\cr
\nonumber\\
&=&P_L \sigma^3
(i{{\cal D}\kern-0.15em\raise0.17ex\llap{/}\kern0.15em\relax})^\dagger
\frac{1}{m^2_s 
+ (i{{\cal D}\kern-0.15em\raise0.17ex\llap{/}\kern0.15em\relax})
(i{{\cal D}\kern-0.15em\raise0.17ex\llap{/}\kern0.15em\relax})^{\dagger}}
\delta(x-y).
\nonumber\\
\end{eqnarray}

\bigskip
\section*{IV. REGULARIZED CURRENTS}
\bigskip

The original gauge current is
\begin{equation}
J^{\mu a} =
\frac{\delta {\cal S}_F}{\delta W_{\mu a}} =
{\tilde{\overline \Psi}}_{L_0}\gamma^\mu{\cal T}^a
\frac{1}{2}(1-\sigma^3){\tilde\Psi}_{L_0}.
\end{equation}
With the regulators, the classical current coupled to $W_{\mu{a}}$ is
\begin{eqnarray}
J^{\mu a}& = &{\tilde{\overline\Psi}}_{L_0}\gamma^\mu{\cal T}^a
\frac{(1-\sigma^3)}{2}{\tilde\Psi}_{L_0} +
\sum_{r= 2,4,...}{\tilde{\overline \Psi}}_{L_r}\gamma^\mu{\cal T}^a
{\tilde\Psi}_{L_r}\cr
\nonumber\\
&+&\sum_{s=1,3,...}{\tilde{\overline \Phi}}_{L_s}\gamma^\mu{\cal T}^a
{\tilde\Phi}_{L_s}.
\end{eqnarray}
Note that there is a $\frac{1}{2}(1-\sigma^3)$ projection associated with 
the undoubled original fermion multiplet. As with conventional 
Pauli-Villars-Gupta regularization, the regulated composite current operator 
is summarized by\cite{Fujinew}
\begin{eqnarray}
\langle J^{\mu a}(x) \rangle_{reg}& = 
&\lim_{x \rightarrow y}Tr\{
-\gamma^\mu(x){\cal T}^a[
\frac{1}{2}(1-\sigma^3)
\langle T\{{\tilde\Psi}_{L_0}(x){\tilde{\overline\Psi}}_{L_0}(y)\}\rangle\cr 
\nonumber\\
&+&\sum_{r=2,4,...}
\langle T\{{\tilde\Psi}_{L_r}(x){\tilde{\overline\Psi}}_{L_r}(y)\}\rangle
+\sigma^3\sum_{s=1,3,...}
\langle T\{{\tilde\Phi}_{L_s}(x){\tilde{\overline\Phi}}_{L_s}(y)\} \rangle
]\}.
\nonumber\\
\end{eqnarray}
The trace runs over Dirac and Yang-Mills indices. 

With the expressions for the propagators and the choice of
$m_n = n\Lambda$ for the regulator masses, we obtain
\begin{eqnarray}
\langle J^{\mu a}(x) \rangle_{reg}&= 
&\lim_{x \rightarrow y}Tr\{
\gamma^\mu(x){\cal T}^aP_L[
\frac{1}{2}(1-\sigma^3)
(i{{\cal D}\kern-0.15em\raise0.17ex\llap{/}\kern0.15em\relax})^{\dagger}
\frac{1}{(i{{\cal D}\kern-0.15em\raise0.17ex\llap{/}\kern0.15em\relax})
(i{{\cal D}\kern-0.15em\raise0.17ex\llap{/}\kern0.15em\relax})^{\dagger}} +\cr 
\nonumber\\
&+&\sum_{r=2,4,...}
(i{{\cal D}\kern-0.15em\raise0.17ex\llap{/}\kern0.15em\relax})^{\dagger}
\frac{1}{r^2\Lambda^2 + 
(i{{\cal D}\kern-0.15em\raise0.17ex\llap{/}\kern0.15em\relax})
(i{{\cal D}\kern-0.15em\raise0.17ex\llap{/}\kern0.15em\relax})^{\dagger}}
-\sum_{s=1,3,...}
(i{{\cal D}\kern-0.15em\raise0.17ex\llap{/}\kern0.15em\relax})^{\dagger}
\frac{1}{s^2\Lambda^2 + 
(i{{\cal D}\kern-0.15em\raise0.17ex\llap{/}\kern0.15em\relax})
(i{{{\cal D}\kern-0.15em\raise0.17ex\llap{/}\kern0.15em\relax}})^\dagger}]
\delta(x-y)\} \cr
\nonumber\\
&=&\lim_{x \rightarrow y}Tr\left\{\gamma^\mu(x)
{\cal T}^a\frac{1}{2}P_L\left[
\frac{1}{i{{\cal D}\kern-0.15em\raise0.17ex\llap{/}\kern0.15em\relax}}
\left(\sum_{n=-\infty}^{\infty}
\frac{(-1)^n{{\cal D}\kern-0.15em\raise0.17ex\llap{/}\kern0.15em\relax}
{{\cal D}\kern-0.15em\raise0.17ex\llap{/}\kern0.15em\relax}^\dagger}
{n^2\Lambda^2 + {{\cal D}\kern-0.15em\raise0.17ex\llap{/}\kern0.15em\relax}
{{\cal D}\kern-0.15em\raise0.17ex\llap{/}\kern0.15em\relax}^{\dagger}}
-\sigma^3\right)\right]\delta(x-y)\right\}\cr
\nonumber\\
&\equiv& \lim_{x \rightarrow y}Tr\left\{\gamma^\mu(x)
{\cal T}^a\frac{1}{2}P_L\left[
\frac{1}{i{{\cal D}\kern-0.15em\raise0.17ex\llap{/}\kern0.15em\relax}}
\left(f({{\cal D}\kern-0.15em\raise0.17ex\llap{/}\kern0.15em\relax}
{{\cal D}\kern-0.15em\raise0.17ex\llap{/}\kern0.15em\relax}^\dagger/\Lambda^2)
-\sigma^3\right)\right]\delta(x-y)\right\}.\cr 
\nonumber\\
\end{eqnarray}
In the above, note that $n$ is summed over {\it all integers}. 

The effect of the tower of regulators is to replace the divergent bare
 expression  
\begin{equation}
\langle J^{\mu a} \rangle_{bare} =
\lim_{x \rightarrow y}Tr\left\{\gamma^\mu(x)
{\cal T}^aP_L\left[
\frac{1}{i{{\cal D}\kern-0.15em\raise0.17ex\llap{/}\kern0.15em\relax}}
\frac{1}{2}\left(1-\sigma^3\right)\right]
\delta(x-y)\right\},
\end{equation}
by 
\begin{equation}
\langle J^{\mu a} \rangle_{reg}=
 \lim_{x \rightarrow y}Tr\left\{\gamma^\mu(x)
{\cal T}^a\frac{1}{2}P_L\left[
\frac{1}{i{{\cal D}\kern-0.15em\raise0.17ex\llap{/}\kern0.15em\relax}}
\left(f({{\cal D}\kern-0.15em\raise0.17ex\llap{/}\kern0.15em\relax}
{{\cal D}\kern-0.15em\raise0.17ex\llap{/}\kern0.15em\relax}^\dagger/\Lambda^2)
-\sigma^3\right)\right]\delta(x-y)\right\}.
\end{equation}
This general feature of the effect of the tower shows up in all the 
regularized gauge currents.

The regulator function
\begin{eqnarray}
f(z) & \equiv & 
\sum_{n=-\infty}^{\infty}\frac{(-1)^n z}{n^2 + z} \cr
\nonumber\\
&=& \frac{\pi\sqrt{z}}{\sinh(\pi\sqrt{z})}
\end{eqnarray}
has the required properties\cite{FS, Okuyama, Fujinew} to ensure convergence. 
For instance, it falls rapidly to zero as $z \rightarrow \infty$, and when 
the regulator masses are taken to $\infty, f(0)=1$. However, the
$\sigma^3$ part of the current remains unmodified, essentially 
 because the tower consists of regulators which are doubled in internal 
space and is  ``$\sigma^3$ neutral''.  It, therefore, can regularize 
only the singlet part of the $\frac{1}{2}(1-\sigma^3)$ projection of the 
bare current. Thus, it remains to be checked 
that for chiral theories free of perturbative anomalies, the 
$\sigma^3$ part gives rise to no further divergences, and can in fact be 
argued to be convergent. When this is true, the total current is then 
successfully regularized by the tower of regulators.

As an example, to specialize to the original Frolov-Slavnov 
proposal\cite{FS} for SO(10), we take
the $d=16$-dimensional representation of Spin(10), 
and set $\Psi^-_L = \Psi_{16}, 
\Psi^+_L = C_{10}\Psi_{\overline{16}}$, and 
$\sigma^3 = \Gamma_{11}$. Note that under a gauge transformation,
\begin{equation}
\Psi_{16} \rightarrow 
exp\left(i\frac{\alpha_a}{2}\Sigma^a(1 -\Gamma_{11})\right)\Psi_{16},
\quad 
C_{10}\Psi_{\overline{16}} \rightarrow 
exp\left(-i\frac{\alpha_a}{2}\left[\Sigma^a(1-\Gamma_{11})\right]^*\right)
C_{10}\Psi_{\overline{16}}
\end{equation}
 due to
$C_{10}\Sigma^a C^{-1}_{10} = -(\Sigma^a)^T = -(\Sigma^a)^*$ and
$\left\{ C_{10},\Gamma_{11}\right\}= 0.$ Here $\Sigma^a$ denote the
generators of SO(10) in the 32-dimensional representation.  
Frolov and Slavnov first observed that for the SO(10) theory with a 
16-dimensional Spin(10) chiral multiplet, it can be argued that the 
$\Gamma_{11}$ part of the current gives rise to no divergences in 
four dimensions since by power counting, the amplitudes for the relevant 
divergent fermion loop diagrams are proportional to the traces of four or 
fewer generators of Spin(10) with $\Gamma_{11}$. These traces vanish 
identically. For arbitrary anomaly-free representations, we shall 
next argue in 
the same spirit that the analogous $\sigma^3$ part of the currents gives rise 
to no further divergences.     

\bigskip
\section*{V. ANOMALIES AND CONDITIONS FOR REGULARIZATION}
\bigskip

We shall expand the vierbein as 
$E^\mu_A = \delta^\mu_A + h^\mu_A$ and 
the rest of the fields about zero.
Then
\begin{eqnarray}
i{{\cal D}\kern-0.15em\raise0.17ex\llap{/}\kern0.15em\relax} &=& 
e^{1\over2}{i{D\kern-0.15em\raise0.17ex\llap{/}\kern0.15em\relax}}
e^{-{1\over2}} \cr
\nonumber\\
&\equiv & i{\partial\kern-0.15em\raise0.17ex\llap{/}\kern0.15em\relax}_f + 
{B\kern-0.15em\raise0.17ex\llap{/}\kern0.15em\relax} 
\end{eqnarray}
where
\begin{equation} 
i{\partial\kern-0.15em\raise0.17ex\llap{/}\kern0.15em\relax}_f = i\gamma^A
\delta^\mu_A\partial_\mu
\end{equation}
is the flat spacetime Dirac operator in the absence of all gauge fields, and
\begin{eqnarray}
{B\kern-0.15em\raise0.17ex\llap{/}\kern0.15em\relax} &=& 
\gamma^A\left[ h^\mu_A i\partial_\mu + (\delta^\mu_A+ h^\mu_A)
(\frac{i}{2}A_{\mu{AB}}\sigma^{AB} -\frac{i}{2}\Gamma^\nu_{\nu\mu}
+ W_{\mu a}{\cal T}^a )\right], \cr 
\nonumber\\
\Gamma^\nu_{\nu\mu} &\equiv& \partial_\mu(ln e).
\end{eqnarray}

The inverse operator has the expansion
\begin{equation}
\frac{1}{i{{\cal D}\kern-0.15em\raise0.17ex\llap{/}\kern0.15em\relax}} = 
\frac{1}{i{\partial\kern-0.15em\raise0.17ex\llap{/}\kern0.15em\relax}}_f + 
\frac{1}{i{\partial\kern-0.15em\raise0.17ex\llap{/}\kern0.15em\relax}}_f
(-{B\kern-0.15em\raise0.17ex\llap{/}\kern0.15em\relax})\frac{1}
{i{\partial\kern-0.15em\raise0.17ex\llap{/}\kern0.15em\relax}}_f +
\frac{1}{i{\partial\kern-0.15em\raise0.17ex\llap{/}\kern0.15em\relax}}_f
(-{B\kern-0.15em\raise0.17ex\llap{/}\kern0.15em\relax})\frac{1}
{i{\partial\kern-0.15em\raise0.17ex\llap{/}\kern0.15em\relax}}_f
(-{B\kern-0.15em\raise0.17ex\llap{/}\kern0.15em\relax})
\frac{1}{i{\partial\kern-0.15em\raise0.17ex\llap{/}\kern0.15em\relax}}_f + ...
\end{equation}
which can be substituted into expression (32). Fermion loops or 
perturbative multipoint correlation functions can be generated by 
functionally differentiating the regularized currents with respect to the 
fields.

In order to obtain the conditions that guarantee no divergences from 
 the unregulated $\sigma^3$ part of (32), we may note that
$Tr(\sigma^3)=0$ and ${B\kern-0.15em\raise0.17ex\llap{/}\kern0.15em\relax}$ 
is linear in ${\cal T}^a$. 
The only term containing ${\cal T}^a$ is $E^\mu_AW_{\mu a}{\cal T}^a$. 
Furthermore, nonvanishing diagrams from the unregulated $\sigma^3$ part 
must involve 
\begin{equation}
Tr\left\{\sigma^3{\cal T}^{a_1}...{\cal T}^{a_n}\right\} \neq 0
\end{equation} 
for some value of $n$. Here, the trace is over internal indices. 

To obtain the condition for convergence, we can consider a generic fermion 
loop diagram with nonvanishing amplitude involving this condition. 
The vertices can be separated into those which involve ${\cal T}^a$ and 
those which do not.
Let $n$ and $k$ be the number of these vertices respectively. 
Potential complications due to curved spacetimes come essentially from 
the ${h^\mu_A}$ terms. Contributions from the Lorentz connection are 
rather straightforward since the coupling involves no 
derivatives and there is also no coupling between $ A_{\mu{AB}}$ 
and ${\cal T}^a$. 
 Diagrams involving vertices due to the
${\overline \Psi}_L{h^\mu_A}i\gamma^A\partial_\mu\Psi_L$ terms 
have derivative couplings. 
These vertices carry weight $\gamma^Ai\partial_\mu$ and diverge linearly.
Thus, the most divergent 
fermion loop diagrams obeying Eq. (39) involve, in general, up to $k$ of these
 $i\gamma^A\partial_\mu$ vertices,
$m$ vertices due to couplings of the type
${\overline \Psi}_L{h^\mu_A}i\gamma^AW_{\mu a}{\cal T}^a\Psi_L$,
and $(n-m)$ vertices
 from  ${\overline \Psi}_Li\gamma^A\delta^\mu_AW_{\mu a}{\cal T}^a\Psi_L$ 
couplings.\footnote{ The number of the two type of vertices 
involving ${\cal T}^a$ must sum up to $n$ to be compatible with Eq. (39).} 
Note also that each propagator between vertices costs 
${{\partial\kern-0.15em\raise0.17ex\llap{/}\kern0.15em\relax}_f}^{-1}$. 
For the purpose of power counting such fermion loops, 
the amplitude therefore behaves symbolically like
\begin{equation}
 \sim Tr\left[\sigma^3 {\cal T}^{a_1}...{\cal T}^{a_n}\right] 
\left(\partial^k \frac{1}{{{\partial\kern-0.15em\raise0.17ex\llap{/}
\kern0.15em\relax}}_f^{m+(n-m)+k}}\right). 
\end{equation}
In momentum space in four dimensions, on integrating over the loop
momentum, this goes like
\begin{equation}
\sim \int d^4p \frac{1}{p^{n}}
Tr\left\{\sigma^3{\cal T}^{a_1}...{\cal T}^{a_n}\right\}.  
\end{equation}
Clearly, the degree of divergence is $(4-n)$. Therefore, if we demand that
\begin{equation}
Tr\left\{\sigma^3{\cal T}^{a_1}...{\cal T}^{a_n}\right\} = 0 \quad {\rm for} 
 \,\, n \leq 4 ,
\end{equation}
 we can argue, as Frolov and Slavnov did for SO(10), that 
the $\sigma^3$ part of the current gives rise to no further divergences, 
and the whole expression (32) is indeed regularized by the tower of regulators.
The condition (42) also suggests generalizations to spacetime
dimensions other than four. 
In terms of  the $T^a$ representation of the original multiplet, Eq. (42) 
translates into
\begin{equation}
Tr\left\{T^{a_1}...T^{a_n}\right\}= 
Tr\left\{(-T^{a_1})^*...(-T^{a_n})^*\right\}, \qquad n \leq 4.
\end{equation}

For real and pseudodreal representations, where $(-T^a)^* = ST^aS^{-1}$, 
the condition is obviously satisfied. This means that all such representations 
are free of perturbative gauge anomalies since the proposed 
regularization which explicitly preserves the gauge symmetries works.
Although it is true that with a different set of regulator masses it may be
 possible to regularize a theory with real representation by a finite tower
instead of the infinite one we have presented, we nevertheless would like to 
discuss the issue in
a more general context.  In this fashion, we will 
obtain a general condition and 
regularization scheme for arbitrary representations including both complex 
and real and pseudoreal representations\cite{Okuyama}. 
However, it may be worthwhile 
to note that for real and pseudoreal representations, the condition
\begin{equation}
Tr\left\{\sigma^3{\cal T}^{a_1}...{\cal T}^{a_n}\right\} = 0 
\end{equation}
is satisfied for {\it all} values of $n$. Thus the $\sigma^3$ contribution of 
the current can be argued to be absent even for {\it convergent} diagrams.
Therefore one can argue that the regularized current is
\begin{equation}
\langle J^{\mu a} \rangle_{reg}=
 \lim_{x \rightarrow y}Tr\left\{\gamma^\mu(x)
{\cal T}^a\frac{1}{2}P_L
\frac{1}{i{{\cal D}\kern-0.15em\raise0.17ex\llap{/}\kern0.15em\relax}}f
({{\cal D}\kern-0.15em\raise0.17ex\llap{/}\kern0.15em\relax}
{{\cal D}\kern-0.15em\raise0.17ex\llap{/}\kern0.15em\relax}^\dagger/\Lambda^2)
\delta(x-y)\right\}.
\end{equation}
A more careful consideration with different choices of the regulator masses
 shows that it is in fact possible to write the regularized currents for real 
and pseudoreal representations in the above form with the appropriate
 regulator functions\cite{Okuyama}.
   
We shall now present the complete solution for arbitrary representations
including complex ones, and show that the invariant 
regularization scheme works if and only if the chiral theory is free of all 
perturbative gauge anomalies, including the mixed Lorentz-gauge anomaly.
 
To begin, note that since the generators 
are Hermitian, $(T^a)^* = (T^a)^T$, and condition (43) is the same as
\begin{eqnarray}
Tr\left\{T^{a_1}...T^{a_n}\right\}& =& 
(-1)^nTr\left\{(T^{a_1})^T...(T^{a_n})^T\right\}\cr
\nonumber\\
&=&(-1)^nTr\left\{(T^{a_n})...(T^{a_1})\right\}, \qquad n \leq 4.
\end{eqnarray}
For $n=2$, the equation is trivially statisfied due to the cyclic property of 
the trace, and this imposes no constraints on $T^a.$
The cases $n=1$ and $n=3$ translate precisely into 
\begin{equation}
Tr(T^a)= 0, \label{eq:one}
\end{equation}
and
\begin{equation}
Tr(T^a\{T^b,T^c\})=0 \label{eq:three}
\end{equation}
respectively. 

The case of $n=4$ imposes no new restrictions if the condition 
for $n=3$ is satisfied. To see this, we decompose the $n=4$ condition into 
antisymmetric and symmetric parts by writing
\begin{eqnarray}
Tr\left(\sigma^3{\cal T}^{a}{\cal T}^{b}{\cal T}^{c}{\cal T}^{d}\right)&=& 
\frac{1}{2}Tr\left(\sigma^3[{\cal T}^{a},{\cal T}^{b}]{\cal T}^c{\cal T}^d
\right)
+\frac{1}{2}Tr\left(\sigma^3\{{\cal T}^{a},{\cal T}^{b}\}{\cal T}^c{\cal T}^d
\right) \cr
\nonumber\\
&=& \frac{i}{2}f^{ab}\,_eTr\left(\sigma^3{\cal T}^e{\cal T}^c{\cal T}^d\right)
+\frac{1}{4}Tr\left(\sigma^3\{{\cal T}^{a},{\cal T}^{b}\}
[{\cal T}^c,{\cal T}^d]\right) \cr
\nonumber\\
&+&\frac{1}{4}Tr\left(\sigma^3\{{\cal T}^{a},{\cal T}^{b}\}
\{{\cal T}^c,{\cal T}^d\}\right).
\end{eqnarray}
The first two of these terms vanish due to 
$[{\cal T}^a, {\cal T}^b]= if^{ab}\,_c{\cal T}^c$, and the $n=3$ 
condition
\begin{equation}
Tr\left(\sigma^3{\cal T}^{a}{\cal T}^b{\cal T}^{c}\right) = 0. 
\end{equation}  
The final term also vanishes, since, in terms of $T^a$,
\begin{eqnarray}
Tr\left(\sigma^3\{{\cal T}^{a},{\cal T}^{b}\}
\{{\cal T}^c,{\cal T}^d\}\right) &=&
Tr\left(\{T^a,T^b\}\{T^c,T^d\}\right)-
Tr\left(\{(T^a)^*,(T^b)^*\}\{(T^c)^*,(T^d)^*\}\right) \cr
\nonumber\\
&=&Tr\left(\{T^a,T^b\}\{T^c,T^d\}\right)-
Tr\left(\{(T^a)^T,(T^b)^T\}\{(T^c)^T,(T^d)^T\}\right) \cr
\nonumber\\
&=&Tr\left(\{T^a,T^b\}\{T^c,T^d\}\right)-
Tr\left(\{T^c,T^d\}\{T^a,T^b\}\right)^T \cr
\nonumber\\
&=&0.
\end{eqnarray}
Thus the constraints on $T^a$ come only from the $n=1$ and $n=3$ restrictions.
The first traceless constraint is precisely the condition in four dimensions
 for the cancellation of the mixed Lorentz-gauge anomaly\cite{nieh}, 
while the second is the requirement for the cancellation of perturbative 
chiral gauge anomalies\cite{Glash}. Therefore, we can conclude that the 
success of this 
invariant scheme is synonymous with the absence of all perturbative gauge 
anomalies. 
In this scheme, it is clear that anomalous chiral theories manifest 
themselves by having unregularized divergent fermion loops, whereas 
anomaly-free theories are regularized in an explicitly invariant manner. 

\bigskip
\section*{VI. GRAVITATIONAL CURRENTS}
\bigskip

We emphasize that in this proposed regularization scheme, all the fermions are
left-handed, and these are coupled to only the left handed projection of 
the spin connection rather than the full spin connection. 
In the chiral representation, 
\begin{equation}
\frac{i}{2}A_{\mu AB}\sigma^{AB}P_L 
= \left[\matrix{0 & 0\cr 0 & A^-_{\mu a}\frac{\tau^a}{2}}\right].
\end {equation}
Note that 
\begin{eqnarray}
A^-_{\mu a}\frac{\tau^a}{2} &=& \left[iA_{\mu 0a}- 
\frac{1}{2}\epsilon_a\,^{bc}A_{\mu bc}\right]\frac{\tau^a}{2} \cr
\nonumber\\
&=& -\frac{i}{4}A_{\mu AB}{\overline\tau}^A\tau^B
\end{eqnarray}
is also precisely the Ashtekar connection in the (anti-)self-dual 
formulation of gravity in four dimensions\cite{ash, samuel}. 
In this context, only 
left-handed fermions are allowed\cite{ART, cps}. Thus 
the regularization scheme studied here is well suited for the 
description of all the four known forces in a completely 
chiral fashion\cite{cps}.   In what follows, we will suppose,
as in the first order formulation, that
the spin connection and the vierbein are independent, and that the 
torsion is not necessarily zero. 

We can compute the the current coupled to the spin 
connection, $A_{\mu AB}$, by using the method of the previous sections. 
In effect, the currents are obtained by the replacement of 
${\cal T}^a$ by
$\frac{i}{2}\sigma^{AB}$ in expressions (28) to (30). The results are 
\begin{equation}
J^{\mu AB}=
{\tilde{\overline \Psi}}_{L_0}\gamma^\mu\frac{i}{2}\sigma^{AB}P_L
\frac{1}{2}(1-\sigma^3){\tilde\Psi}_{L_0}.
\end{equation}
and
\begin{eqnarray}
\langle J^{\mu AB}(x) \rangle_{reg}&=&\lim_{x \rightarrow y}Tr\{
\gamma^\mu(x)\frac{i}{2}\sigma^{AB}P_L[
\frac{1}{2}(1-\sigma^3)(i{{\cal D}\kern-0.15em\raise0.17ex\llap{/}
\kern0.15em\relax})^{\dagger}
\frac{1}{(i{{\cal D}\kern-0.15em\raise0.17ex\llap{/}\kern0.15em\relax})
(i{{\cal D}\kern-0.15em\raise0.17ex\llap{/}\kern0.15em\relax})^\dagger} \cr 
\nonumber\\
&+&\sum_{r=2,4,...}
(i{{\cal D}\kern-0.15em\raise0.17ex\llap{/}\kern0.15em\relax})^{\dagger}
\frac{1}{r^2\Lambda^2 + 
(i{{\cal D}\kern-0.15em\raise0.17ex\llap{/}\kern0.15em\relax})
(i{{\cal D}\kern-0.15em\raise0.17ex\llap{/}\kern0.15em\relax})^{\dagger}}
-\sum_{s=1,3,...}
(i{{\cal D}\kern-0.15em\raise0.17ex\llap{/}\kern0.15em\relax})^{\dagger}
\frac{1}{s^2\Lambda^2 + 
(i{{\cal D}\kern-0.15em\raise0.17ex\llap{/}\kern0.15em\relax})
(i{{\cal D}\kern-0.15em\raise0.17ex\llap{/}\kern0.15em\relax})^{\dagger}}
]\delta(x-y)\} \cr
\nonumber\\
&=& \lim_{x \rightarrow y}Tr\left\{\gamma^\mu(x)
{\frac{i}{2}\sigma^{AB}}\frac{1}{2}P_L\left[
{\frac{1}{i{{\cal D}\kern-0.15em\raise0.17ex\llap{/}\kern0.15em\relax}}}
\left(f({{\cal D}\kern-0.15em\raise0.17ex\llap{/}\kern0.15em\relax}
{{\cal D}\kern-0.15em\raise0.17ex\llap{/}\kern0.15em\relax}^\dagger/\Lambda^2)
-\sigma^3\right)\right]\delta(x-y)\right\}.\cr 
\end{eqnarray} 

Again, the presence of the tower of regulators serves to replace the 
divergent bare
 expression  
\begin{equation}
\langle J^{\mu AB} \rangle_{bare} =
\lim_{x \rightarrow y}Tr\left\{\gamma^\mu(x)
{\frac{i}{2}\sigma^{AB}}P_L\left[
\frac{1}{i{{\cal D}\kern-0.15em\raise0.17ex\llap{/}\kern0.15em\relax}}
\frac{1}{2}\left(1-\sigma^3\right)\right]
\delta(x-y)\right\}
\end{equation}
by 
\begin{equation}
\langle J^{\mu AB} \rangle_{reg}=
 \lim_{x \rightarrow y}Tr\left\{\gamma^\mu(x)
{\frac{i}{2}\sigma^{AB}}\frac{1}{2}P_L\left[
\frac{1}{i{{\cal D}\kern-0.15em\raise0.17ex\llap{/}\kern0.15em\relax}}
\left(f({{\cal D}\kern-0.15em\raise0.17ex\llap{/}\kern0.15em\relax}
{{\cal D}\kern-0.15em\raise0.17ex\llap{/}\kern0.15em\relax}^\dagger/\Lambda^2)
-\sigma^3\right)\right]\delta(x-y)\right\}.
\end{equation}
The arguments of the last section with regard to the 
unregularized $\sigma^3$ part can
be repeated, and 
we conclude that Eqs. (47) and (48) are again the 
precise conditions for the current to be free of divergences.

Next, we discuss the energy-momentum tensor. Various proposals for 
defining the 
 energy momentum tensor have been suggested\cite{Fujienergy}.
If the classical bare action is regarded as
$S_F\left({\overline\Psi_{L_0}}, \Psi_{L_0}, E^\mu_A, W_{\mu a}, A_{\mu AB}
\right)$,
then the energy momentum tensor $\Theta_{\mu\nu}$ is obtained 
from  
\begin{eqnarray}
e\Theta_{\mu\nu}&=& e_{\mu A}\frac{\delta S_F}{\delta E^\nu_A}\cr
\nonumber\\
&=&{\overline\Psi_{L_0}}{\gamma_\mu}iD_\nu\Psi_{L_0} -g_{\mu\nu}{\cal L},
\end{eqnarray}
where ${\cal L}$ is the Lagrangian.
On the other hand, if the variables 
${\tilde{\overline\Psi}_L}$ and ${\tilde\Psi}_L$ are to be treated as 
independent integration variables as is suggested by the
 diffeomorphism-invariant measure (15), 
then the energy momentum tensor $T_{\mu\nu}$
regarded as the source current for the background vierbein is 
\begin{equation}
eT_{\mu\nu}= e_{\mu A}\frac{\delta{\tilde S}_F}{\delta E^\nu_A}
\end{equation}
with 
\begin{eqnarray}
{\tilde S}_F\left({\tilde{\overline\Psi}_{L_0}}, {\tilde\Psi}_{L_0}, 
E^\mu_A, W_{\mu a}, A_{\mu AB}\right) &=&
\int d^4x {\tilde{\overline\Psi}}_{L_0}e^{1\over2}i
{{D\kern-0.15em\raise0.17ex\llap{/}\kern0.15em\relax}}e^{-{1\over2}}
{\tilde\Psi}_{L_0} \cr
\nonumber\\
&=&\int d^4x {\tilde{\overline\Psi}}_{L_0}E^\mu_A\gamma^A
\left[iD_\mu -\frac{i}{2}(\partial_\mu \ln e)\right]\tilde\Psi_{L_0}.
\nonumber\\
\end{eqnarray}
The expression for the corresponding energy-momentum tensor is then
\begin{equation}
eT_{\mu\nu}=
{\tilde{\overline\Psi}_{L_0}}{\gamma_\mu}
i\left(D_\nu -\frac{1}{2}\Gamma^\alpha_{\alpha\nu}\right){\tilde\Psi}_{L_0}
-\frac{i}{2}g_{\mu\nu}
\partial_\alpha({\tilde{\overline\Psi}}_{L_0}\gamma^\alpha{\tilde\Psi_{L_0}}).
\nonumber\\
\end{equation}
In terms of variables which are not densitized,
\begin{equation}
T_{\mu\nu}={{\overline\Psi}_{L_0}}{\gamma_\mu}iD_\nu{\Psi}_{L_0}
-\frac{i}{2}g_{\mu\nu}\left[\partial_\alpha(
{{\overline\Psi}}_{L_0}\gamma^\alpha{\Psi}_{L_0})+
\Gamma^\beta_{\beta\alpha}{{\overline\Psi}}_{L_0}\gamma^\alpha{\Psi}_{L_0})
\right].
\end{equation}
As a result, $T_{\mu\nu}$ and $\Theta_{\mu\nu}$ are related by
\begin{equation}
eT_{\mu\nu} = e\Theta_{\mu\nu} + \frac{1}{2}g_{\mu\nu}
\left({\tilde{\overline\Psi}}_{L_0}\frac{\delta{\tilde S}_F}
{\delta{\tilde{\overline\Psi}}_{L_0}} + {\tilde\Psi}_{L_0}
\frac{\delta{\tilde S}_F}{\delta{\tilde\Psi}_{L_0}}\right).
\end{equation}
The difference between the two is therefore not significant classically 
when the equations of motion can be imposed. However, at the quantum level, 
there can be subtleties\cite{Fujienergy}. 
Because of the choice of the densitized variables, all bare mass terms and, in 
particular, regulator mass terms, are independent of the vierbein and 
therefore do {\it not} contribute to $T_{\mu\nu}$. The total energy-momentum
tensor will include the kinetic, but no mass, contributions of all the 
anticommuting and commuting regulators. The regularized expression becomes
\begin{eqnarray}
\langle eT_{\mu\nu}\rangle_{reg}&=&
\lim_{x \rightarrow y}Tr\left\{{\gamma_\mu}
i\left(D_\nu -\frac{1}{2}\Gamma^\alpha_{\alpha\nu}\right)
\frac{1}{i{{\cal D}\kern-0.15em\raise0.17ex\llap{/}\kern0.15em\relax}}P_L 
\frac{1}{2}
\left[f\left(\frac{{{\cal D}\kern-0.15em\raise0.17ex\llap{/}\kern0.15em\relax}
{{\cal D}\kern-0.15em\raise0.17ex\llap{/}\kern0.15em\relax}^\dagger}
{\Lambda^2}\right)
-{\sigma^3}\right]\delta(x-y)
\right\} \cr
\nonumber\\
&+&\frac{i}{2}g_{\mu\nu}\langle\partial_\alpha J^\alpha_5 \rangle, 
\label{eq:regtmn}
\end{eqnarray}
where $J^\alpha_5$ is the ABJ current which will be discussed more fully in
the next section:
\begin{equation}
{J^\mu_5}= -\sum_{r=0,2,...}{\tilde{\overline\Psi}}_{L_r}
\gamma^\mu\tilde\Psi_{L_r}
+\sum_{s=1,3,...}{\tilde{\overline\Phi}}_{L_s}\sigma^3
\gamma^\mu\tilde\Phi_{L_s}.\nonumber
\end{equation}
Again, a slight variation of the previous arguments with 
regard to relevant diagrams proves that the $\sigma^3$ part of 
the energy-momentum tensor gives rise to no divergent fermion loops if 
conditions (47) and (48) hold. Hence the 
expression for the energy-momentum tensor is regularized for finite $\Lambda$.

In our present discussion, we do not densitize the 
{\it background} variables and eschew use, for instance, of
$W_{Aa}\equiv e^{1\over2}E^\mu_AW_{\mu a}$ instead of $W_{\mu a}$. This choice
would be useful if an explicitly diffeomorphism invariant measure 
$\prod DW_{Aa}$ is required when the path integral formalism is to be 
applied to the quantization of the gauge fields\cite{Fujienergy}. In this 
paper, gauge and gravitational fields are to be treated as background 
fields only. 

The energy-momentum tensor
should be symmetrized if it is to be regarded as the source of the metric. 
It is known that there are no perturbative Lorentz anomalies in four 
dimensions\cite{Chang}. This is {\it verified by the explicitly 
Lorentz-invariant 
regularization scheme} proposed here. 
If the vierbein and the left-handed
spin connection are to be dynamically described by the 
(anti-)self-dual formulation of gravity\cite{ash, samuel}, then the 
energy momentum tensor appears as the 
source on the right hand side of the corresponding equation of motion.
We also do not Hermitize the Weyl action. The difference between 
the Weyl action and the Hermitian version involves the divergence
of the ABJ current and also torsion terms and, is given by Eq. (36) of Ref.
\cite{cps}. As a result, among other things, the energy-momentum tensor 
presented here picks up an imaginary term (in Lorentzian signature 
spacetimes) proportional to the divergence of the chiral current. 
Since the expectation value of the divergence of the ABJ current is not 
zero quantum mechanically, there can be subtle violations of
discrete symmetries due to the ABJ anomaly\cite{ABJ}, especially in the 
presence of topologically nontrivial gauge and gravitational instantons, 
and also nonvanishing torsion.  Details of
consequences of these violations will be presented elsewhere.

\bigskip
\section*{VII. $\gamma^5$ ANOMALY}
\bigskip

The regularization of gauge singlet currents requires a separate discussion. 
The ABJ current has aleady appeared above
in the regularized expression for the energy-momentum tensor 
Eq. (\ref{eq:regtmn}) and, as
we shall see, plays a critical role in constraining the fermion content
of the theory\cite{cps}.  

Under a singlet chiral $\gamma^5$ rotation, 
\begin{eqnarray}
{\tilde\Psi}_{L_r} \rightarrow 
e^{i\alpha\gamma^5}{\tilde\Psi}_{L_r} &=& e^{-i\alpha}{\tilde\Psi}_{L_r},\cr
\nonumber\\  
{\tilde{\overline\Psi}}_{L_r} \rightarrow 
{\tilde{\overline\Psi}}_{L_r} e^{i\alpha\gamma^5}
&=&{\tilde{\overline\Psi}}_{L_r}e^{i\alpha}, \label{eq:abj}
\end{eqnarray}
and similarly for ${\tilde\Phi}_{L_s}$ and ${\tilde{\overline\Phi}}_{L_s}$.
Kinetic terms are invariant under this global tranformation, but mass terms 
are not.
The bare massless action is invariant under such a global transformation, and
 the associated ABJ or $\gamma^5$ current 
\begin{equation} 
J^\mu_5 = 
{\tilde{\overline\Psi}}_{L_0}\gamma^\mu\gamma^5\tilde\Psi_{L_0}
= -{\tilde{\overline\Psi}}_{L_0}\gamma^\mu\tilde\Psi_{L_0}
=-J^\mu_F,
\end{equation}
is conserved classically, i.e.. $\partial_\mu J^\mu_5 = 0.$
However, the bare quantum composite current 
\begin{equation}
\langle J^{\mu}_5 \rangle_{bare} =
-\lim_{x \rightarrow y}Tr\left\{\gamma^\mu(x)P_L\left[
\frac{1}{i{{\cal D}\kern-0.15em\raise0.17ex\llap{/}\kern0.15em\relax}}\frac{1}
{2}\left(1-\sigma^3\right)\right]
\delta(x-y)\right\}
\end{equation}
is divergent. The regularized current is not necessarily conserved. 
In the generalized Pauli-Villars-Gupta scheme, the mass terms of the 
regulators break the symmetry explicitly. So for the ABJ current, even at the 
classical level, the current including the regulators is only partially 
conserved. The relation is
\begin{eqnarray}
\partial_\mu {J^\mu_5} =&& i[\sum_{r=2,4,...}
m_r({\tilde\Psi}^T_{L_r}\sigma^1C_4{\tilde\Psi}_{L_r} 
- {\tilde{\overline\Psi}}_{L_r}{C^\dagger_4}
\sigma^1{\tilde{\overline\Psi}}^T_{L_r})\cr
\nonumber\\
&-&{\sum_{s=1,3,...}}m_s({\tilde\Phi}^T_{L_s}\sigma^1
\sigma^3C_4{\tilde\Phi}_{L_s} - 
{\tilde{\overline\Phi}}_{L_s}C^\dagger_4\sigma^3\sigma^1
{\tilde{\overline\Phi}}^T_{L_s})],
\nonumber\\
\end{eqnarray}
with
\begin{equation}
{J^\mu_5}= -\sum_{r=0,2,...}{\tilde{\overline\Psi}}_{L_r}
\gamma^\mu\tilde\Psi_{L_r}
+\sum_{s=1,3,...}{\tilde{\overline\Phi}}_{L_s}\sigma^3
\gamma^\mu\tilde\Phi_{L_s}.
\end{equation}
The expectation value of the regularized ABJ current is
\begin{eqnarray}
\langle J^{\mu}_5(x) \rangle_{reg}&=&-\lim_{x \rightarrow y}Tr\{
\gamma^\mu(x)\frac{i}{2}P_L[
\frac{1}{2}(1-\sigma^3)(i{{\cal D}\kern-0.15em\raise0.17ex\llap{/}
\kern0.15em\relax})^{\dagger}
\frac{1}{(i{{\cal D}\kern-0.15em\raise0.17ex\llap{/}\kern0.15em\relax})
(i{{\cal D}\kern-0.15em\raise0.17ex\llap{/}\kern0.15em\relax})^\dagger} \cr 
\nonumber\\
&+&\sum_{r=2,4,...}(i{{\cal D}\kern-0.15em\raise0.17ex\llap{/}
\kern0.15em\relax})^{\dagger}
\frac{1}{r^2\Lambda^2 + 
(i{{\cal D}\kern-0.15em\raise0.17ex\llap{/}\kern0.15em\relax})
(i{{\cal D}\kern-0.15em\raise0.17ex\llap{/}\kern0.15em\relax})^{\dagger}}
-\sum_{s=1,3,...}
(i{{\cal D}\kern-0.15em\raise0.17ex\llap{/}\kern0.15em\relax})^{\dagger}
\frac{1}{s^2\Lambda^2 + 
(i{{\cal D}\kern-0.15em\raise0.17ex\llap{/}\kern0.15em\relax})
(i{{\cal D}\kern-0.15em\raise0.17ex\llap{/}\kern0.15em\relax})^{\dagger}}
]\delta(x-y)\} \cr
\nonumber\\
&=&-\lim_{x \rightarrow y}Tr\left\{\gamma^\mu(x)
\frac{1}{2}(1-\gamma^5)
{\frac{1}{i{{\cal D}\kern-0.15em\raise0.17ex\llap{/}\kern0.15em\relax}}}
\frac{1}{2}\left(f({{\cal D}\kern-0.15em\raise0.17ex\llap{/}\kern0.15em\relax}
{{\cal D}\kern-0.15em\raise0.17ex\llap{/}\kern0.15em\relax}^\dagger/\Lambda^2)
-\sigma^3\right)\delta(x-y)\right\}.\cr   \label{eq:abjreg}
\end{eqnarray} 

The previous arguments concerning the unregulated $\sigma^3$ 
part are still valid.  Within this context, we have in effect 
regularized  the 
ABJ current,
and the associated amplitudes 
can be computed explicitly.

There is nevertheless a subtlety which we have glossed over.  The 
transformation given by Eq. (\ref{eq:abj}) rotates fields by the same phase,
independently of their quantum numbers under the gauge group.  That 
is the reason why mass terms are not left invariant. We could
suppose, on the other hand, that
the phase transformation on ${\tilde\Psi}_{L_0}$ is generated by 
fermion number and consider the regulator fields to consist of fermions 
and antifermions.
In this fashion, the relevant fermion number current can be written as
\begin{equation}
{J^\mu_{f}}= \sum_{r=0,2,...}{\tilde{\overline\Psi}}_{L_r}
\sigma^3\gamma^\mu\tilde\Psi_{L_r}
+\sum_{s=1,3,...}{\tilde{\overline\Phi}}_{L_s}
\gamma^\mu\tilde\Phi_{L_s}. \label{eq:vect}
\end{equation}
This current should be conserved classically.  However,
a straightforward repeat of the arguments above now shows that it is
not regularized.  The factor of $(1-\sigma^3)$ that appears in the primary
field ${\tilde\Psi_{L_0}}$ is modified upon regularization to $(1-f\sigma^3)$.
As a result, regularization affects the part of the amplitude which was
convergent because of Eqs. (\ref{eq:one}) and (\ref{eq:three}), but leaves 
the remainder divergent.  

It is interesting to note that a similar phenomenon takes place in the
 vectorlike formulation\cite{Fujinew}. If we double in external rather than 
internal space by including bispinors regulators, then the corresponding 
result for the fermion current is 
\begin{equation}
\langle J^{\mu}_f \rangle_{reg}=
- \lim_{x \rightarrow y}Tr\left\{\gamma^\mu(x)
\frac{1}{2}\left(f({{\cal D}\kern-0.15em\raise0.17ex\llap{/}\kern0.15em\relax}
{{\cal D}\kern-0.15em\raise0.17ex\llap{/}\kern0.15em\relax}^\dagger/\Lambda^2)
-\gamma^5\right)\frac{1}{i{{\cal D}\kern-0.15em\raise0.17ex\llap{/}
\kern0.15em\relax}}\frac{1}{2}(1-\sigma^3)\delta(x-y)\right\}.
\end{equation} 
In this scheme,\footnote{In the vectorlike formulation, the doubling is in 
external space, and the covariant derivative in Eq. (73) contains 
$W_{\mu a}T^a$ rather than $W_{\mu a}{\cal T}^a$.} potentially 
unregularized divergences can 
come from the $\gamma^5(1-\sigma^3)$ part of $J^{\mu}_f$.
For {\it gauge} currents, parity-odd divergent contributions from
fermion loops cancel for anomaly-free gauge theories.
However, for $J^\mu_f$, parity-odd amplitudes from triangle diagrams remain 
unaffected by these restrictions and are divergent\cite{Fujinew}. 
For the axial current, the regularized
chiral projection operator appears as $(1 -f\gamma_5)$,
and there are potential divergences in the parity-conserving part of the 
amplitude. Nonetheless, it can be shown that the divergent diagrams cancel when
the anomaly cancellation conditions hold\cite{Okuyama}.

In the chiral scheme, it is possible to define the fermion current via
the ABJ current, $J^\mu_F= -J^\mu_5$, with both currents carrying weight 
1 as a result of the choice of the densitized commuting and 
anticommuting variables. This
identification is consistent with the original degeneracy present
in the bare action. Note that the {\it chiral} regularization preserves this 
degeneracy\footnote{See also Eqs. (66) and (67).}. 
No such definition of regularized singlet currents free of 
divergences is possible within the vectorlike scheme without further 
auxilliary regularization prescriptions\cite{Fujinew}.

The ABJ anomaly 
can be explicitly computed by taking the divergence 
of the expectation value of the regularized expression (\ref{eq:abjreg}). 
Here we choose to
compute the explicit divergence of the chiral current as
\begin{equation}
\langle \partial_\mu J^\mu_5 \rangle_{reg} =
\partial_\mu \lim_{x \rightarrow y}Tr\left\{-\gamma^\mu
\frac{1}{2}(1-\gamma^5)
{\frac{1}{i{{\cal D}\kern-0.15em\raise0.17ex\llap{/}\kern0.15em\relax}}}
\frac{1}{2}\left(f({{\cal D}\kern-0.15em\raise0.17ex\llap{/}\kern0.15em\relax}
{{\cal D}\kern-0.15em\raise0.17ex\llap{/}\kern0.15em\relax}^\dagger/\Lambda^2)
-\sigma^3\right)\delta(x-y)\right\}. 
\end{equation} 
The trace can be evaluated by using the complete sets of 
eigenvectors, $\{X_n\}$ and $\{Y_n\}$, of the 
positive-semidefinite Hermitian operators with
\begin{eqnarray}
{{\cal D}\kern-0.15em\raise0.17ex\llap{/}\kern0.15em\relax}
{{\cal D}\kern-0.15em\raise0.17ex\llap{/}\kern0.15em\relax}^\dagger X_n &=& 
\lambda^2_n X_n, \cr
\nonumber\\ 
{{\cal D}\kern-0.15em\raise0.17ex\llap{/}\kern0.15em\relax}^\dagger
{{\cal D}\kern-0.15em\raise0.17ex\llap{/}\kern0.15em\relax} Y_n &=& 
\lambda^2_n Y_n.
\end{eqnarray}
For the modes with nonzero eigenvalues, $X_n$ and $Y_n$ are 
paired by\footnote{It is assumed that zero modes
have been subtracted from the expectation value of the current. They 
do not occur in the action in the path integral 
formulation\cite{Fujikawa}.}   
\begin{equation}
X_n = {{\cal D}\kern-0.15em\raise0.17ex\llap{/}\kern0.15em\relax} Y_n/
\lambda_n, \qquad Y_n ={{\cal D}\kern-0.15em\raise0.17ex\llap{/}
\kern0.15em\relax}^\dagger{X_n}/ \lambda_n.
\end{equation}
Consequently, this yields
\begin{eqnarray}
\langle \partial_\mu J^\mu_5 \rangle_{reg} &=&
-\partial_\mu\left[ \sum_n X^{\dagger}_n{\gamma^\mu} P_L
(i{{\cal D}\kern-0.15em\raise0.17ex\llap{/}\kern0.15em\relax})^\dagger
{\frac{1}{{{\cal D}\kern-0.15em\raise0.17ex\llap{/}\kern0.15em\relax}
{{\cal D}\kern-0.15em\raise0.17ex\llap{/}\kern0.15em\relax}^\dagger}}
\frac{1}{2}
\left(f({{\cal D}\kern-0.15em\raise0.17ex\llap{/}\kern0.15em\relax}
{{\cal D}\kern-0.15em\raise0.17ex\llap{/}\kern0.15em\relax}^\dagger/\Lambda^2)
-\sigma^3\right)X_n\right] \cr
\nonumber\\
&=&
i\partial_\mu \left[\sum_n X^\dagger_n{\gamma^\mu}P_L
\frac{1}{2\lambda_n}\left(f(\lambda^2_n/\Lambda^2)
-\sigma^3\right)Y_n \right]\cr
\nonumber\\
&=&i[\sum_n \partial_\mu (X^\dagger_n\gamma^\mu)P_L
\frac{1}{2\lambda_n}(f(\lambda^2_n/\Lambda^2) -\sigma^3)Y_n \cr
\nonumber\\
&+&\sum_n X^\dagger_nP_R\frac{1}{2\lambda_n}(f(\lambda^2_n/\Lambda^2)
-\sigma^3)\gamma^\mu\partial_\mu Y_n ] \cr
\nonumber\\
&=&-\frac{i}{2}\sum_n[Y^\dagger_n\frac{1}{2}(1-\gamma^5)
(f({{\cal D}\kern-0.15em\raise0.17ex\llap{/}\kern0.15em\relax}^\dagger
{{\cal D}\kern-0.15em\raise0.17ex\llap{/}\kern0.15em\relax}/\Lambda^2)
-\sigma^3)Y_n \cr
\nonumber\\
&-& X^\dagger_n\frac{1}{2}(1+\gamma^5)
(f({{\cal D}\kern-0.15em\raise0.17ex\llap{/}\kern0.15em\relax}
{{\cal D}\kern-0.15em\raise0.17ex\llap{/}\kern0.15em\relax}^\dagger/\Lambda^2)
-\sigma^3)X_n].
\nonumber\\ \label{eq:abjanomaly}
\end{eqnarray}  
The traces over $\sigma^3$ as well as the parity-even part drop out. On 
taking the limit of infinite regulator 
masses($\Lambda \rightarrow \infty$), the result for Euclidean signature is
\begin{eqnarray}
\langle \partial_\mu J^\mu_5 \rangle_{reg} 
&=&\lim_{\Lambda \rightarrow \infty}
\frac{i}{4}\sum_n[Y^\dagger_n\gamma^5f
({{\cal D}\kern-0.15em\raise0.17ex\llap{/}\kern0.15em\relax}^\dagger
{{\cal D}\kern-0.15em\raise0.17ex\llap{/}\kern0.15em\relax}/\Lambda^2)Y_n
+ X^\dagger_n\gamma^5f
({{\cal D}\kern-0.15em\raise0.17ex\llap{/}\kern0.15em\relax}
{{\cal D}\kern-0.15em\raise0.17ex\llap{/}\kern0.15em\relax}^\dagger/
\Lambda^2)X_n] \cr
\nonumber\\
&=& {{i \times d}\over{768\pi^2}}F_{\alpha\beta AB}
\epsilon^{\alpha\beta\mu\nu}F_{\mu\nu}\,^{AB}
+{i\over{32\pi^2}}Tr(\epsilon^{\alpha\beta\mu\nu}
G_{\alpha\beta a}T^aG_{\mu\nu b}T^b).
\nonumber\\
\end{eqnarray}  
$G_{\mu\nu a}$ and $F_{\mu\nu AB}$ are, respectively, the curvatures of
$W_{\mu a}$ and $A_{\mu AB}$.
Note that in the first line of Eq. (78) there is a factor of 
${1\over 4}$ in the 
trace over 2$d$-dimensional internal space, and Dirac indices. 
This gives the result which is {\it one-half} of the chiral anomaly of a 
vector theory. Because all the fields are Weyl, the factor we get for
the gravitational part is also $d$ rather than 2$d$. This is in agreement with
the fact that there are $d$ Weyl fermions coupled to gravity in the bare 
action.

\newpage
\bigskip
\section*{VIII. REMARKS}
\bigskip

We have presented a generalization of the Frolov-Slavnov invariant  
regularization  scheme for chiral fermion theories in curved spacetimes. 
The Lagrangian level
regularization is explicitly invariant under all the local gauge symmetries of 
the theory, including local Lorentz invariance. The perturbative scheme works 
if and only if the chiral gauge anomaly and the 
mixed Lorentz-gauge anomaly cancellation conditions hold. Anomalous theories 
manifest themselves in having divergent fermion loops which remain
unregularized by the scheme. 
Since the invariant scheme is promoted to include local Lorentz invariance, 
spectator fields which do not couple to gravity cannot be, and are not,
introduced. Furthermore, in the proposed scheme, the theory is truly 
chiral (Weyl) in that all fields are left-handed, including the regulators,
and only the left-handed spin connection is needed. The scheme is therefore 
well suited for the study of the interaction of matter with all the four 
known forces in a completely chiral manner. 
In contrast with the 
vector-like formulation, the degeneracy between the ABJ current 
and the fermion number current in the bare action is preserved by the 
regularization.

How would nonperturbative effects such as global
anomalies appear in the scheme?
As presented, the scheme is perturbative and
the success of the scheme is predicated upon the absence of perturbative 
gauge anomalies. A general discussion on nonperturbative effects is 
outside of the scope
of this paper.  Instead, we will focus on
two ways these effects can be recognized, together with one
significant consequence.   For instance,
it is clear that the 
perturbative scheme regularizes a theory with a single left-handed internal 
SU(2) doublet. Yet, it is known that such a theory suffers from 
the SU(2) global anomaly\cite{Witten, Alwis}.  
By embedding SU(2) in SU(3), the gauge SU(2) global anomaly is shown 
to be related to the perturbative SU(3) chiral gauge anomaly\cite{Nair}. 
Within the present context, there are then
fermion loops containing SU(3) vertices which fail to be regularized. 
Anomalies also manifest themselves in 
path integrals as nontrivial Jacobians in the 
measure under a change of variables\cite{Fujikawa}.  From this perspective,
the global SU(2) anomaly gives rise to an inconsistent Jacobian when the 
transformation of ($-$1) is 
considered both as a $2\pi$ rotation in SU(2) and as a $\pi$ rotation 
induced by $\gamma^5$
in nontrivial $\theta$ vacua\cite{cps, Alwis}.  The present regularization 
scheme will not control all divergent amplitudes in these
sectors.  
Thus a path integral
formulation dependent on the 
tower of regulators may yield further consistency conditions from
cancellation of nontrivial Jacobians.\footnote{In this respect, 
the situation 
may be clearer in a truly non-perturbative formulation such as the 
overlap formalism for the fermion determinant\cite{Nara}.}
As an example, if a Euclidean path integral is to include all topologies 
for four manifolds\cite{Hawking} 
and hence the required generalized spin structures\cite{hp}, then a 
further global Lorentz anomaly cancellation condition selects grand unified 
theories with multiples of 16 Weyl fermions\cite{cps, cfg}. 
 
Finally, it may be worthwhile to calculate the effective action generated
by the theory. For instance, it is known\cite{Birrell} that the 
Einstein-Hilbert-Palatini action and the cosmological term are among the 
counterterms when a fermion is quantized
in background curved spacetime with parity conservation. 
The explicitly {\it chiral}-invariant 
regularization scheme presented here may be used to check the resultant
requisite
counter terms with parity nonconservation. It is possible for example, 
if the torsion is not assumed to vanish, that the 
Samuel-Jacobson-Smolin\cite{samuel} action of 
the (anti-)self-dual formulation of gravity may 
emerge instead from integration
over the fermion and regulator fields.

\bigskip\bigskip\bigskip
\section*{ACKNOWLEDGMENTS}
\bigskip

The research for this work has been supported by the Department of Energy
under Grant No. DE-FG05-92ER40709-A005.

\end{document}